\documentclass[%
 reprint,
 amsmath,amssymb,
 aps,
]{revtex4-1}
\usepackage{slashed}
\usepackage{graphicx}
\usepackage{dcolumn}
\usepackage{bm}

\begin{document}

DMUS--MP--16/17\\
\\
\title{Massless sector of $AdS_3$ superstrings: a geometric interpretation}
\author{Andrea Fontanella}
\email{email: \tt a.fontanella@surrey.ac.uk} \author{Alessandro Torrielli}%
 \email{email: \tt a.torrielli@surrey.ac.uk}
\affiliation{%
 \vspace{2mm} Department of Mathematics, University of Surrey,
\\ Guildford, GU2 7XH, UK\\}%





\begin{abstract}
We study the recently discovered $q$-deformed Poincar\'e supersymmetry of the $AdS_3/CFT_2$ integrable massless scattering, and demonstrate how the $S$-matrix is invariant under boosts. The boost generator has a non-local coproduct, which acts on the scattering matrix as a differential operator, annihilating it. We propose to reinterpret the boost action in terms of covariant derivatives on bundles, and derive an expression for the $S$-matrix as the path-ordered exponential of a flat connection. We provide a list of possible alternative interpretations of this emergent geometric picture, including a one-dimensional auxiliary Schr\"odinger problem. We support our claims by performing a simplified algebraic Bethe ansatz, which bears some resemblance to antiferromagnets.   
 
\end{abstract}

\maketitle


\section{\label{sec:level1}Introduction}




Integrability in the AdS/CFT correspondence \cite{Beisert:2010jr,Arutyunov:2009ga} is a source of ever new mathematical structures in the theory of exact S-matrices. This is well demonstrated by superstrings on $AdS_3\times S^3\times S^3\times S^1$ and $AdS_3\times S^3\times T^4$ \cite{Babichenko:2009dk,rev3}. The  superisometry algebra of the first background is the $\mathfrak{D}(2,1;\alpha)\times \mathfrak{D}(2,1;\alpha)$ Lie superalgebra, with $\alpha$ measuring the relative radius of the $S^3$'s. The second background is obtained by In\"on\"u-Wigner contraction of the first as $\alpha \to 0$, and has superisometry $\mathfrak{psu}(1,1|2)\times\mathfrak{psu}(1,1|2)$. 

Classical integrability was proven in \cite{Babichenko:2009dk,Sundin:2012gc}, and the finite-gap equations were given in \cite{OhlssonSax:2011ms}. The massive scattering problem was solved in \cite{Borsato:2012ud,Borsato:2012ss,
Borsato:2013qpa,Borsato:2013hoa} based on residual centrally-extended  $\mathfrak{psu}(1|1)$ symmetry factors, with satisfactory matching against perturbative string predictions \cite{Rughoonauth:2012qd,Sundin:2012gc,
Abbott:2012dd,Beccaria:2012kb,
Beccaria:2012pm,Sundin:2013ypa,
Bianchi:2013nra}. However, the novel appearance of {\it massless modes}  \cite{Sax:2012jv,Lloyd:2013wza} complicates the world-sheet analysis \cite{Borsato:2014exa,Borsato:2014hja,
Abbott:2014rca}. The massless sector was recently unfolded in \cite{Sax:2014mea,Borsato:2016kbm, upcom}, and a natural field-theory dual candidate shown to exactly align to the group-theoretical procedure of \cite{Sax:2012jv}, although mismatches with perturbation theory \cite{PerLinus} are not entirely resolved. Further work is found in \cite{Abbott:2013ixa,Sundin:2013uca,
Borsato:2015mma,Prin,Abbott:2015mla,Per,
Pittelli:2014ria,Regelskis:2015xxa}. 

\subsection{The $q$-deformed Poincar\'e superalgebra} 

When considering the $AdS_5$ scattering problem, the authors of \cite{Gomez:2007zr,Charles} interpreted the massive magnon dispersion relation as the Casimir of a $q$-deformed Poincar\'e superalgebra. This algebra was never quite turned into a full symmetry of the $S$-matrix, but a boost generator was introduced and used to re-obtain the uniformising rapidity torus \cite{Beisert:2010jr,Arutyunov:2009ga}. Further appearance of $q$-Poincar\'e symmetries occurred {\it e.g.} in \cite{Pachol:2015mfa,Kenta}. 

In \cite{Joakim} it was shown that such a deformed algebra is an {\it exact symmetry} of the massless $AdS_3$ $S$-matrix. Combined with the massless dispersion relation, the new structure has quite natural features, providing for example a rather compact formulation of the comultiplication rule. Some interesting analogies with phonon physics, inspired by \cite{Ballesteros:1999ew}, are also noticeable in this setup. The boost generator was then used in \cite{Joakim} to derive a particular uniformising rapidity, reproducing the standard Zamolodchikov's massless variable in the relativistic limit. 

\subsection{This paper}

In the present paper, we demonstrate precisely in which sense the $S$-matrix is invariant under the boost-generator coproduct. The boost acts on the scattering matrix as a differential operator, while the coproduct's {\it tail} multiplies it as a matrix. The combination of the two actions annihilates the $S$-matrix. This complements \cite{Joakim}, hence one concludes that the whole $q$-Poincar\'e superalgebra is a symmetry of the massless $AdS_3$ $S$-matrix.

The boost symmetry bears more resemblance to a geometric condition - the $S$-matrix being covariantly constant with respect to a connection - than to a purely algebraic constraint. We therefore attempt to reinterpret geometrically the invariance equations in terms of covariant differentiation on bundles, which leads us to re-express the very $S$-matrix as the path-ordered exponential of a flat connection. This connection admits several singular points on the base space, which is reminiscent of Aharonov-Bohm type problems. This is not a new prerogative of quantum group $R$-matrices, and in Appendix B we refer to closely related classic literature.

We then give a list of possible alternative pictures, including a reinterpretation of the $S$-matrix as the propagator in a one-particle auxiliary quantum mechanics, where a parametrisation of the original scattering momenta is translated into the {\it time} variable. The Hamiltonian of this subsidiary problem is then related to a simplified algebraic Bethe ansatz, which singles out free fermions with gapless dispersion relation, resembling the one of Heisenberg antiferromagnets. We believe such analogies deserve further exploration, which we plan for future work.



\section{\label{summary}Massless $R$-matrix and boosts}

In this section, we summarise the results of \cite{Joakim} which will be needed for our present purposes. 

The centrally-extended $\mathfrak{su}(1|1)_L \oplus \mathfrak{su}(1|1)_R $ Lie superalgebra reads
\begin{eqnarray}
\label{alge}
&&\{\mathfrak{Q}_{L}, \mathfrak{S}_{L}\} = \mathfrak{H}_{L}, \quad \{\mathfrak{Q}_{R}, \mathfrak{S}_{R}\} = \mathfrak{H}_{R}, \nonumber \\
&&\{\mathfrak{Q}_{L}, \mathfrak{Q}_{R}\} = \mathfrak{P}, \qquad  \{\mathfrak{S}_{L}, \mathfrak{S}_{R}\} = \mathfrak{K}. 
\end{eqnarray}
The r.h.s. of (\ref{alge}) displays four central elements. One represents (\ref{alge}) on a boson-fermion doublet $\{|\phi\rangle, |\psi\rangle\}$:
\begin{eqnarray}
\label{leftrep}
&&\mathfrak{S}_{R} = - \sqrt{h \sin \frac{p}{2}}\begin{pmatrix}0&0\\1&0\end{pmatrix},\quad
\mathfrak{Q}_{R} = - \sqrt{h \sin \frac{p}{2}}\begin{pmatrix}0&1\\0&0\end{pmatrix}, \nonumber \\\nonumber
&&\mathfrak{Q}_{L} = \sqrt{h \sin \frac{p}{2}}\begin{pmatrix}0&0\\1&0\end{pmatrix},\quad
\mathfrak{S}_{L} = \sqrt{h \sin \frac{p}{2}}\begin{pmatrix}0&1\\0&0\end{pmatrix}, \\
&&\qquad \qquad \mathfrak{H}_{L} = \mathfrak{H}_{R} = - \mathfrak{P} = - \mathfrak{K} = h \, \sin \frac{p}{2}.
\end{eqnarray}
The central element $p$ is the {\it momentum}, with physical eigenvalues in $[0,2 \pi]$, while $\mathfrak{H} = \mathfrak{H}_L + \mathfrak{H}_R$ is the {\it energy}. The $R$-matrix which was studied, namely 
\begin{equation}\label{eq:RLL}
  \begin{aligned}
    &R |\phi\rangle \otimes |\phi\rangle\ = {} \, |\phi\rangle \otimes |\phi\rangle, \\
    &R |\phi\rangle \otimes |\psi\rangle\ = - {} \, \csc \frac{p_1 + p_2}{4} \, \sin \frac{p_1 - p_2}{4} |\phi\rangle \otimes |\psi\rangle + \\
    &\qquad \qquad {} \, \csc \frac{p_1 + p_2}{4} \, \sqrt{\sin \frac{p_1}{2} \sin \frac{p_2}{2}} |\psi\rangle \otimes |\phi\rangle, \\
    &R |\psi\rangle \otimes |\phi\rangle\ = {} \, \csc \frac{p_1 + p_2}{4} \, \sin \frac{p_1 - p_2}{4}  |\psi\rangle \otimes |\phi\rangle + \\
    &\qquad \qquad {} \, \csc \frac{p_1 + p_2}{4} \, \sqrt{\sin \frac{p_1}{2} \sin \frac{p_2}{2}} |\phi\rangle \otimes |\psi\rangle, \\
    &R |\psi\rangle \otimes |\psi\rangle\ = - {} \, |\psi\rangle \otimes |\psi\rangle,\nonumber
\end{aligned}
\end{equation}
satisfies the equation
\begin{equation}\label{definiLL}
  \Delta_N^{\text{op}} (\mathfrak{a})\,  R\ =\ R\, \Delta_N (\mathfrak{a}) \, \, \, \, \forall \, \, \mathfrak{a} \in \mathfrak{su}(1|1)_{L}\oplus \mathfrak{su}(1|1)_{R},
\end{equation}
where $\Delta_N^{op} = \Pi (\Delta_N)$, $\Pi$ being the graded permutation on the tensor-product algebra $\Pi (\mathfrak{a} \otimes \mathfrak{b}) = (-)^{|\mathfrak{a}| |\mathfrak{b}|} \mathfrak{b} \otimes \mathfrak{a}$, and 
\begin{equation}
  \begin{aligned}
     &\Delta_N(p) = \, p \otimes \mathfrak{1} + \mathfrak{1} \otimes p, \\
     &\Delta_N(\mathfrak{P})= \, \mathfrak{P} \otimes e^{i \frac{p}{2}} + e^{-i \frac{p}{2}} \otimes \mathfrak{P}, \\
    &\Delta_N(\mathfrak{K})= \, \mathfrak{K} \otimes e^{i \frac{p}{2}} + e^{-i \frac{p}{2}} \otimes \mathfrak{K},  \\
      &\Delta_N(\mathfrak{H}_{R})= \, \mathfrak{H}_{R} \otimes {e^{i \frac{p}{2}}} + {e^{-i \frac{p}{2}}} \otimes \mathfrak{H}_{R}, \\
      &\Delta_N(\mathfrak{H}_{L})= \, \mathfrak{H}_{L} \otimes {e^{i \frac{p}{2}}} + {e^{-i \frac{p}{2}}} \otimes \mathfrak{H}_{L},\\\nonumber
 \end{aligned}
\end{equation}
\begin{equation}
  \label{coprod}
  \begin{aligned}
        &\Delta_N(\mathfrak{Q}_{L}) = \, \mathfrak{Q}_{L} \otimes e^{i \frac{p}{4}} + e^{-i \frac{p}{4}} \otimes \mathfrak{Q}_{L}, \\
     &\Delta_N(\mathfrak{S}_{L}) = \, \mathfrak{S}_{L} \otimes e^{i \frac{p}{4}} + e^{-i \frac{p}{4}} \otimes \mathfrak{S}_{L}~,\\
     &\Delta_N(\mathfrak{Q}_{R}) = \, \mathfrak{Q}_{R} \otimes {e^{i \frac{p}{4}}} + {e^{-i \frac{p}{4}}} \otimes \mathfrak{Q}_{R}, \\
     &\Delta_N(\mathfrak{S}_{R})= \, \mathfrak{S}_{R} \otimes {e^{i \frac{p}{4}}} + {e^{-i \frac{p}{4}}} \otimes \mathfrak{S}_{R}.  
  \end{aligned}
\end{equation}
Formulas (\ref{coprod}) prescribe how the symmetry algebra acts on two-particle states, in such a way that it is a representation of (\ref{alge}). $R$ satisfies the Yang-Baxter equation 
\begin{eqnarray}
\label{YBE}
&&R_{12}(p_1,p_2) \, R_{13}(p_1,p_3) \,R_{23}(p_2,p_3) \, \nonumber\\
&&\qquad \qquad  = R_{23}(p_2,p_3) \,R_{13}(p_1,p_3) \,R_{12}(p_1,p_2),
\end{eqnarray}
defined on the {\it triple} tensor-product space, {\it e.g.}
\begin{equation}
R_{12} \, |\phi\rangle \otimes |\phi\rangle \otimes |\phi\rangle = R \otimes \mathfrak{1} \, |\phi\rangle \otimes |\phi\rangle \otimes |\phi\rangle, \qquad etc.,  
\end{equation}
and the condition of so-called {\it braiding unitarity}, {\it i.e.} $\Pi(R)(p_2,p_1) \, R(p_1,p_2) = \mathfrak{1} \otimes \mathfrak{1}$. The $R$-matrix also satisfies $\Pi(R)(p_2,p_1) = R(p_1,p_2)$. To describe the scattering of massless $AdS_3$ modes, $R$ needs to be multiplied by the scalar factor, which we call $\Phi$, and was calculated in \cite{upcom}. So equipped, and up to a permutation of the outgoing particles, $R$ is the $S$-matrix, scattering particle $1$ - with 1D momentum $p_1$ - with particle $2$ - with momentum $p_2$.

In \cite{Joakim}, these conditions were used to describe two copies of the 1+1 dimensional {\it $q$-deformed super-Poincar\'e algebra}, denoted by $\mathfrak{E}_q(1,1)_{L} \oplus \mathfrak{E}_q(1,1)_{R}$. 
The coupling constant $h$ is related to the deformation parameter by
\begin{equation}
\log q \equiv \frac{i}{h^2}.
\end{equation}
We define $\mu \equiv \frac{4}{h^2}$, and the relations (\ref{alge}) are extended to
\begin{eqnarray}
\label{algeq}
&&\{\mathfrak{Q}_{R}, \mathfrak{S}_{R}\} \ = \ \mathfrak{H}_{R}, \quad \{\mathfrak{Q}_{L}, \mathfrak{S}_{L}\} \ = \ \mathfrak{H}_{L}, \quad [\mathfrak{J}_{R}, p] \ = \ i \mathfrak{H}_{R}, \nonumber \\
&& [\mathfrak{J}_{L}, p] \ = \ i \mathfrak{H}_{L},\nonumber \quad [\mathfrak{J}_A, \mathfrak{H}_B] \ = \frac{e^{i p} - e^{-i p}}{2 \mu},\nonumber \\
&& [\mathfrak{J}_A, \mathfrak{Q}_B] \ = \frac{i}{2 \sqrt{\mu}} \frac{e^{i \frac{p}{2}} + e^{- i \frac{p}{2}}}{2} \mathfrak{Q}_B, \nonumber \\
&&  [\mathfrak{J}_A, \mathfrak{S}_B] = \frac{i}{2 \sqrt{\mu}} \frac{e^{i \frac{p}{2}} + e^{- i \frac{p}{2}}}{2} \, \mathfrak{S}_B,
\end{eqnarray}
where {\footnotesize $(A,B) = (L,L), (L,R), (R,L), (R,R)$}, and
\begin{eqnarray}
\label{algeqc}
&&\{\mathfrak{Q}_{L}, \mathfrak{Q}_{R}\} \ = \ \mathfrak{P}~, \qquad  \{\mathfrak{S}_{L}, \mathfrak{S}_{R}\} \ = \ \mathfrak{K},\nonumber \\ 
&&[\mathfrak{J}_{L}, \mathfrak{P}] \ = [\mathfrak{J}_{R}, \mathfrak{P}] \ = \ [\mathfrak{J}_{L}, \mathfrak{K}] \ = [\mathfrak{J}_{R}, \mathfrak{K}]= \frac{e^{- i p} - e^{i p}}{2 \mu}.\nonumber
\end{eqnarray}
The (suitably normalised) quadratic Casimir is given by  
\begin{eqnarray}
\mathfrak{C}_2 \equiv \mathfrak{H}^2  - 4 h^2 \sin \frac{p}{2}. \nonumber
\end{eqnarray}
The massless representation is characterised by the vanishing of the Casimir (massless {\it dispersion relation}).

The coproduct for the boost operator, say, $\mathfrak{J}_L$, compatible with the algebra relations, reads (cf. \cite{Charles})
\begin{eqnarray}
\label{deltaJ}
&&\Delta_N (\mathfrak{J}_L) = \mathfrak{J}_L \otimes e^{i \frac{p}{2}} +  e^{-i \frac{p}{2}} \otimes \mathfrak{J}_L +\nonumber\\
&&\quad \, \, \frac{1}{2} \, \mathfrak{Q}_L \, e^{- i \frac{p}{4}} \otimes \mathfrak{S}_L \, e^{i \frac{p}{4}} +  \frac{1}{2} \, \mathfrak{S}_L \, e^{- i \frac{p}{4}} \otimes \mathfrak{Q}_L \, e^{i \frac{p}{4}},
\end{eqnarray}
with a {\it tail} given by a bilinear expression in the supercharges.
The boost operators act as
\begin{equation}
\mathfrak{J}_R = i \mathfrak{H}_R \, \partial_p, \qquad \mathfrak{J}_L = i \mathfrak{H}_L \, \partial_p.
\end{equation}
The form of the boost coproduct (\ref{deltaJ}) suggests that the worldsheet realisation of $\mathfrak{J}$ should be non-local, according to a standard argument reviewed {\it e.g.} in \cite{Beisert:2010jr}-VI.2. 

Without loss of generality, for the rest of the paper we focus on the $L$ part only.

\section{\label{sec:level2}Covariant derivatives}

The symmetry conditions one imposes for the $R$-matrix, invariant under the generators of a given Hopf algebra, is determined by (\ref{definiLL}). 

This applies to all the supercharges and central elements of the $q$-Poincar\'e superalgebra. When it comes to the boost generator $\mathfrak{J}$, which acts as a derivative, the situation is slightly different, although we are still after a constraint on the $R$-matrix which is as close as possible to {\it boost invariance}. By tedious though straightforward calculation, one can verify that the following holds:
\begin{equation}
\label{boost}
\boxed{  
\Delta_N(\mathfrak{J}_L) (R) = 0 = \Delta_N^{op}(\mathfrak{J}_L) (R),}
\end{equation}
(recall that $R$ is normalised to have the all-bosonic entry equal to $1$ - {\it i.e.}, it is stripped off of the scalar factor $\Phi$). 
We therefore consider the {\it action} of the boost coproducts as $\mathfrak{J}$ differentiating the individual $R$-matrix entries, and the tail acting multiplicatively. We have taken states not to transform under boosts, {\it i.e.} $\mathfrak{J} |\phi\rangle =\mathfrak{J} |\psi\rangle = 0$, while algebra generators do transform, in accordance with the Heisenberg picture of quantum mechanics.

Spelling out the coproducts explicitly, one notices a first curious condition implied by the above equations:
\begin{equation}
\label{covder}
D R = 0, \qquad D \equiv \frac{1}{2 h \sin \frac{p_1}{2} \sin \frac{p_2}{2}} \Big[\Delta_N(\mathfrak{J}_L) - \Delta_N^{op}(\mathfrak{J}_L)\Big],\nonumber
\end{equation}
which reads out
\begin{equation}
\label{out}
D = \frac{\partial}{\partial p_2} - \frac{\partial}{\partial p_1} + \Gamma(p_1,p_2) \big[ E_+ \otimes E_- + E_- \otimes E_+\big],\, \, 
\end{equation}
where $\Gamma(p_1,p_2) = \frac{\cos \frac{p_1 - p_2}{4}}{2 \sqrt{\sin \frac{p_1}{2} \sin \frac{p_2}{2}}}$, and 
\begin{equation} 
E_+ \equiv \begin{pmatrix}0&1\\0&0\end{pmatrix}, \qquad E_- \equiv \begin{pmatrix}0&0\\1&0\end{pmatrix}.\nonumber 
\end{equation} 
 
After taking suitable combinations of the two equations in (\ref{boost}), we then obtain the following system:
\begin{equation}
\label{7}
D_M R \equiv \bigg[\frac{\partial}{\partial p_M} + \Gamma_M\bigg]R = 0, \qquad M =1,2,
\end{equation}
\begin{eqnarray}
\label{first}
&& \Gamma_1 \equiv -\frac{1}{4} \sqrt{\frac{\sin \frac{p_2}{2}}{\sin \frac{p_1}{2}}} \, \frac{\big[ E_+ \otimes E_- + E_- \otimes E_+\big]}{\sin \frac{p_1 + p_2}{4}},\nonumber \\
&& \Gamma_2 \equiv \frac{1}{4} \sqrt{\frac{\sin \frac{p_1}{2}}{\sin \frac{p_2}{2}}} \, \frac{\big[ E_+ \otimes E_- + E_- \otimes E_+\big]}{\sin \frac{p_1 + p_2}{4}}.
\end{eqnarray} 
This resembles a covariant derivative on a 2-dimensional manifold $\mathcal{B}$ with {\it real coordinates} $(p_1,p_2)$. Because of the $8\pi$-periodicity in both coordinates, $\mathcal{B}$ is topologically equivalent to $T^2$. The periodicity is four times the physical strip of left- and right-moving massless modes \cite{upcom}. 

We remark that the two equations (\ref{7}) are not independent if one uses $\Pi(R)(p_2,p_1) = R(p_1,p_2)$. In our approach, we propose to regard the two equations (\ref{7}) as fundamental, from which one can derive $\Pi(R)(p_2,p_1) = R(p_1,p_2)$. Braiding unitarity of the $R$-matrix will then be considered as a constraint equation.
 
The above conditions allow us to write an integral formula for the $R$-matrix. 
We take a curve $\gamma(\lambda) : [0, 1] \rightarrow \mathcal{B} \simeq T^2$, and contract (\ref{7}) with $\frac{d p^M}{d \lambda}$. Integrating along $\gamma$ gives the following formal expression for the $R$-matrix:
\begin{equation}
\label{int_R}
R \big[\gamma (\lambda)\big] = \Pi_s\,  e^{ \int_{\gamma(0)}^{\gamma(\lambda)} dp^M \Gamma_M }, 
\end{equation}
where $\Pi_s$ is the graded permutation operator acting on two-particle states as $\Pi (|v\rangle \otimes |w\rangle) = (-)^{|v| |w|} |w\rangle \otimes |v\rangle$, and $\cal{P}$ denotes the path-ordering of the exponential. Notice that the sign in the exponent of (\ref{int_R}) is due to the fact that we have singled out $\Pi_s$ for convenience, moreover
\begin{equation}
\{\Pi_s, \Gamma_M\} = 0,
\end{equation}
hence $\Big[\frac{\partial}{\partial p_M} - \Gamma_M\Big]\Pi_s \circ R = 0$, and of course $\Pi_s^2=\mathfrak{1}\otimes \mathfrak{1}$. 

The starting point of integration must be chosen in accordance with the fact that $R$ reduces to $\Pi_s$ for equal values of the two momenta $p_1=p_2$, which can easily be seen from (\ref{eq:RLL}). In Appendix A we provide an explicit calculation with a choice of contour, for which it is easier to verify that (\ref{int_R}) exactly reproduces the expression (\ref{eq:RLL}). 

If we choose $\gamma$ to be a closed loop and trace (\ref{int_R}) over the superspin states, we obtain what, in gauge-theory language, is the \textit{Wilson loop} associated with $\Gamma_M$. 

We remark that, as a consequence of (\ref{7}) and as can be verified by explicit computation, $\Gamma_M$ is locally flat ({\it pure gauge}), since its curvature $F_{MN}$ is vanishing:
\begin{equation}
\label{flatness}
F_{12} = \partial_1 \Gamma_2 - \partial_2 \Gamma_1 + [\Gamma_1,\Gamma_2] = 0.
\end{equation}
Hence we expect a $g$ exists such that locally $\Gamma = g^{-1} dg$. 

Of course, all this applies away from singularities and for paths entirely contained within regular regions. This reminds of Aharonov-Bohm type effects, and would obstruct the global trivialisation of any associated bundle one were to attach to this description (cf. next section). 

\smallskip

An important remark is that $\Gamma_1$ and $\Gamma_2$ are proportional to the same matrix in $\mathfrak{su}(1|1)_L \otimes \mathfrak{su}(1|1)_L$. Nevertheless, there is a non-trivial structure inherited by the fiber. The matrix defining $\Gamma_M$ has fermionic indices. Moreover, we might think of its specific form as deriving from a dynamical principle, whereby $\Gamma$ itself is a solution of the equations of motions and the Bianchi identity.

\smallskip

Finally, if we multiplied the $R$-matrix by the phase factor $\Phi$, we would still be able to write an equation like
\begin{equation}
\label{fase}
\bigg[\frac{\partial}{\partial p_M} + \Gamma_M - \frac{\partial}{\partial p_M} \log \Phi\bigg]\tilde{R} = 0, \qquad \tilde {R} \equiv \Phi R.
\end{equation}

\section{Geometric interpretation}

As discussed in \cite{Charles}, boost invariance of the $R$-matrix in relativistic integrable systems amounts to 
\begin{eqnarray}
\label{singleo}
(J^{rel} \otimes \mathfrak{1} + \mathfrak{1} \otimes J^{rel}) R = \bigg[\frac{\partial}{\partial \theta_1} + \frac{\partial}{\partial \theta_2}\bigg] R = 0,
\end{eqnarray}
where the operator acting on $R$ is the trivial coproduct of the relativistic boost $J^{rel} = \frac{\partial}{\partial \theta}$, which shifts the particle rapidity $\theta$. Since the coproduct is equal to its opposite, the equation (\ref{singleo}) one obtains from invariance $[\Delta(\mathfrak{J}^{rel}),R]=0$ implies difference-form $R = R(\theta_1 - \theta_2)$.

In our case, as we have motivated earlier, this is altogether more complicated. One can interpret (\ref{boost}) as a deformation of relativistic boost invariance. The purpose of the following discussion is to propose alternative pictures, which might link to entirely independent mathematical structures underlying the problem. It is then fascinating to ask what these various emerging interpretations might bring back to the original physics.

Likewise, it is interesting to imagine what the axioms of integrable scattering (such as {\it crossing symmetry}, the {\it bootstrap principle} and the {\it Yang-Baxter equation}) might become, when seen in the light of these alternative frameworks. For instance, starting with the path-ordered expression (\ref{int_R}), one may try to associate {\it braiding unitarity}
\begin{eqnarray}
\Pi(R)(p_2,p_1) \, R(p_1,p_2) = \mathfrak{1} \otimes \mathfrak{1}
\end{eqnarray}
with path-inversion:
$
e^{ - \int_{\gamma(0)}^{\gamma(\lambda)} dp^M \Gamma_M } \, e^{ \int_{\gamma(0)}^{\gamma(\lambda)} dp^M \Gamma_M } \  = \mathfrak{1} \otimes \mathfrak{1}$.

When searching for an analogue of crossing symmetry, one needs a suitable continuation to complex momenta, and a putative complexified bundle (with a more complicated base manifold). The path $\bar{\gamma}$ integrating (\ref{int_R}) to {\it crossed} regions \cite{upcom} should have the problem of avoiding the singularities of $\Gamma$ on the base space.

The Yang-Baxter equation (\ref{YBE}) would be suggestive of a condition on the holonomy of $\Gamma$ (cf. next subsection).

\subsection{Flat connections on bundles}

The emergence of $\Gamma_M$, which takes values in $\mathfrak{su}(1|1)_L \otimes \mathfrak{su}(1|1)_L$, hints at a description in terms of a bundle, $P(\mathcal{B}, \mathcal{F} ; \pi)$, where the base space $\mathcal{B}$ is topologically equivalent to $T^2$, and the fiber $\mathcal{F}$ a subset of $U(\mathfrak{su}(1|1)_L) \otimes U(\mathfrak{su}(1|1)_L)$, where $U$ denotes the universal enveloping algebra. The $R$-matrix would be a local section of $P$, and a scalar function from base-space perspective. Pointwise, the connection $\Gamma_M$ should uniquely split the tangent space of $P$ into vertical and horizontal subspaces. Adding (\ref{fase}) extends the fiber by a subset of $\mathbb{R}$.

In this language, the Yang-Baxter equation could be interpreted geometrically.
Let us consider the embedding of the torus $T^2$ into the 3D space $(p_1,p_2,p_3)$, where $p_3$ stands for the momentum of the third auxiliary particle. Let us consider three different points $X_1, X_2, X_3\in T^2$ and denote by $\gamma_{ij}$ a path from $X_i$ to $X_j$.  If the holonomy of $\Gamma$ is \emph{trivial}, then the initial and final points of the  horizontal lift of a closed loop on $T^2$ must coincide. This reads in terms of the $R$-matrix as $R_{12}R_{23} = R_{13}$. However this is \emph{not} the Yang-Baxter equation, which points towards the non-triviality of the holonomy of $\Gamma$. 

The flatness of the connection would technically define a homomorphism between the fundamental group of $\mathcal{B}$ and the holonomy group of the connection $\Gamma$. In the trivial case, since the fundamental group of $\mathcal{B}\simeq T^2$ is $\mathbb{Z}\times \mathbb{Z}$, the holonomy of $\Gamma$ would depend by a pair of integers $(n_1, n_2)$ - which are the winding numbers of the loop $\gamma$ on $T^2$ - and not on the shape of the loop $\gamma$, while our paths are of course affected by the singularities of $\Gamma$. 

If we assume to be able to reduce, at least locally, to a principal-bundle structure, then at a given point $p$ in the non-vanishing intersection of two patches $U_i$ and $U_j$
\begin{equation}
\Gamma_{M}^i = t_{ij}^{-1} \Gamma_M^j t_{ij}  + t_{ij}^{-1} \partial_M t_{ij},
\end{equation}
where $t_{ij}$ is the transition function. Let $|p\rangle$ be a local section over $U_i \cap U_j$, such that
\begin{equation}
|p\rangle_i = t_{ij} |p\rangle_j.
\end{equation}
The expression of the connection $\Gamma_1$ on the patch $U_i$ should then be given by
\begin{equation}
\Gamma_1^i (p) =-\frac{1}{4} \sqrt{\frac{\sin \frac{p_2}{2}}{\sin \frac{p_1}{2}}} \, \frac{_i\langle p| \big[ E_+ \otimes E_- + E_- \otimes E_+\big]|p\rangle_i}{\sin \frac{p_1 + p_2}{4}}. \nonumber
\end{equation}
We plan to study the singularities of $\Gamma$ and the fiber structure in future work. 

In this setup, $R$ would be a {\it covariantly constant section} of $P$, and it has the natural interpretation of holonomy of the connection $\Gamma$. It is interesting to speculate that there might exist a {\it gauge-theory} rewriting of our problem, where the gauge field $\Gamma_M$ lives on $T^2$, with gauge algebra given by two copies of $\mathfrak{su}(1|1)$. It would then be curious to investigate what a {\it gauge transformation} might correspond to in the original physical picture. This should turn into a {\it local} - {\it i.e.} momentum-dependent - tranformation of the basis of two-particle scattering states ({\it i.e.} a local redefinition of the Faddeev-Zamolodchikov operators \cite{Arutyunov:2009ga}). Because of its local nature, it should tie in with the $\mathfrak{sl}(2)$ outer-automorphism of $\mathfrak{su}(1|1)_L \oplus \mathfrak{su}(1|1)_R$.

In this thinking scheme, the path-ordered expression (\ref{int_R}) would coincide with a particular {\it Wilson line} of the (almost everywhere flat) gauge connection $\Gamma_M$. 

\section{\label{sec:level22}One-particle Schr\"odinger equation}

Either equation (\ref{7}) can also be interpreted as an auxiliary Schr\"odinger problem:
\begin{equation}
\label{purp}
i \hbar \frac{d}{dt} U(t_0;t) = H U(t_0;t),
\end{equation}
where the role of the final time $t$ is taken by one of the two momenta, the other one being the initial time $t_0$. The path-ordered exponential solution produces the quantum mechanical {\it propagator} $U(t_0;t)$: 
\begin{equation}
R = \Pi_s \circ {\cal{P}} e^{- \frac{i}{\hbar}. \int_{t_0}^t H}
\end{equation}
In general, we would now speak of a {\it trajectory} on the torus $T^2: \{ p^1,p^2 \in [0,8 \pi]\}$, such that, for $\tau \in [0,t]$,
\begin{equation}
p^M = p^M(\tau), \qquad p^M(0) = (p_0,p_0), \qquad p^M(t) = (p_1,p_2),\nonumber
\end{equation}
for some $p_0$, one has
\begin{eqnarray}
&&\frac{dp^M}{d\tau}\bigg[\frac{\partial}{\partial p^M} + \Gamma_M(p^1,p^2)\bigg]R = \bigg[\frac{d}{d\tau} + \dot{p}^M \Gamma_M\bigg]R = 0,\nonumber\\
&&R = \Pi_s {\cal{P}} e^{\int_0^t \, \dot{p}^M \Gamma_M}, \qquad \dot{p}^M \equiv \frac{dp^M}{d\tau}.
\end{eqnarray}
$\cal{P}$ is now a time-ordering. Any trajectory provides an alternative Schr\"odinger problem. The flatness of $\Gamma_M$, and the fact (proved by straightforward computation) that $\dot{p}^M \Gamma_M$ is identically zero along the line $p^1=p^2$, make all these quantum mechanical problems equivalent.

It is not artificial to think of $\frac{dp^M}{d\tau} \Gamma_M$ as a Hamiltonian (although possibly singular). As shown in the next section, its spin structure matches the Hamiltonian-density $h_{n,n+1}$ emerging from a simplified algebraic Bethe ansatz. In fact, (\ref{out}) is consistent with the definition of $h_{n,n+1}$ as a logarithmic derivative of $R$. In (\ref{purp}), however, $H$ does not act on spin-chain sites, but on the superspin degrees of freedom of an auxiliary quantum-mechanical particle. 

Moreover, given the fact that such an auxiliary Hamiltonian arises from the symmetrised tensor-product of two supercharges, it strongly resembles what one has in ${\cal{N}}=1$ supersymmetric quantum mechanics, where the $\mathfrak{su}(1|1)$ supersymmetry algebra determines the energy as
\begin{equation}
H = \{Q,Q^\dagger\}.
\end{equation}  

Further remarks are annotated in Appendix B.

\section{\label{sec:level3}Gapless spin-chains}

In this section, we perform a simplified Bethe ansatz with the purpose of motivating the previous section's Hamiltonian approach. We shall follow \cite{Kulish}. From the $R$-matrix we can obtain a spin-chain Hamiltonian in the following standard fashion. We define
\begin{eqnarray}
p_1 = i \nu + \lambda, \qquad p_2 = i \nu - \lambda,
\end{eqnarray}
and notice that
\begin{eqnarray}
R(\nu, \lambda=0) = \Pi_s.
\end{eqnarray}
We construct a local spin-chain Hamiltonian $\cal{H}$ by setting
\begin{eqnarray}
&&p_n = i \nu_n + \lambda_n, \quad p_{n+1} = i \nu_n - \lambda_n, \quad  \nonumber \\
&&h_{n,n+1} = \frac{\partial}{\partial \lambda_n} \log R(\nu, \lambda_n)_{\vert \lambda_n=0} = \, \Pi_s \circ \frac{\partial}{\partial \lambda_n} R(\nu, 0), \nonumber
\end{eqnarray}
where the chain has $N$ sites and is periodic, {\it i.e.} $N+1 \equiv 1$, and $h_{n,n+1}$ acts on neighbouring sites $(n,n+1)$. 

The simplification consists in taking $\nu_n = \nu \ \forall n$, giving 
\begin{eqnarray}
h_{n,n+1} = \frac{i}{2 \sinh \frac{\nu}{2}} \, (E_+ \otimes E_- + E_- \otimes E_+),
\end{eqnarray}
which is a Hermitean operator for real $\nu$ (taking into account fermionic signs).

{\it Remark.} The Hamiltonian density coincides, up to an overall factor, with the tail of the boost coproduct, equivalently with either of $\Gamma_M$. It is tantalising to consider this a sort of {\it massless Dirac Hamiltonian} on the chain.

The algebraic Bethe ansatz, with monodromy matrix
\begin{equation}
T \equiv L_{01} (\lambda) ... L_{0N} (\lambda),
\end{equation}
built from the Lax matrix $L_{0i}=R_{0i}$, gives a Hamiltonian
$$
{\cal{H}} = \frac{d}{d\lambda} \log tr T (0) = -\sum_{n=1}^N h_{n,n+1}.
$$
Consider now an infinite chain, with pseudo-vacuum \begin{eqnarray}
|\Omega\rangle = |... \phi \otimes \phi \otimes \phi ... \rangle,
\end{eqnarray}
which is an eigenstate of ${\cal{H}}$ with zero energy. The spectrum of one-particle excitations above $|\Omega\rangle$ is {\it gapless}:
\begin{eqnarray}
\label{dis}
&&|\Psi_p\rangle = \sum_m e^{i p m} |... \phi_{m-1} \otimes \psi_m \otimes \phi_{m+1} ... \rangle,\nonumber\\
&&{\cal{H}}|\Psi_p\rangle = \epsilon \sin p \, |\Psi_p\rangle, \qquad \epsilon = \frac{1}{\sinh \frac{\nu}{2}}.
\end{eqnarray}
There are intriguing similarities between the dispersion relation (\ref{dis}) and the one of {\it spinons},  massless excitations of the {\it antiferromagnetic} Heisenberg spin-chain \cite{Faddeev:1981ip}:
\begin{equation}
\label{spinon}
E_{sp} = \frac{\pi}{2} \, |\sin p |.
\end{equation}
Notice that also {\it massless (ferromagnetic) magnons} (\ref{leftrep}) have a dispersion relation $E = 2 h \, |\sin \frac{p}{2} |$ which is very similar to the {\it massless spinons} (\ref{spinon}), and one wonders of any possible spectral duality between the two. As it was done for $AdS_5$, where antiferromagnets appear in specific parameter scalings \cite{H1,H2}, it would be very interesting to study the antiferromagnetic limit of the $AdS_3$ massless sector and test this idea, particularly in view of \cite{PerLinus}.

\section{Conclusions}
In this paper we have presented a geometric interpretation of the scattering of massless modes in $AdS_3$ integrable superstring theory. This involves rephrasing the boost invariance of the $S$-matrix into a parallel condition with respect to a covariant derivative with flat connection, and attempting a dictionary to bundles. The $S$-matrix plays the role of a Wilson line, and physical conditions, such as braiding unitarity and the Yang-Baxter equation, have suggestive interpretations. The connection displays singular points, which might turn into the appearance of Aharonov-Bohm type effects. 

We have then outlined a possible alternative picture in terms of a quantum mechanics with purely spin degrees of freedom, where the scattering momenta are parametrised by a time coordinate. We have substantiated this view with the help of a simplified spin-chain Bethe ansatz, characterised by gapless excitations with dispersion relation bearing similarity to Heisenberg antiferromagnets.

In the appendices we provide a consistency check and some remarks. We speculate that the different interpretations we discussed might converge into a Berry-phase type description, and notice resemblance with the Knizhnik-Zamolodchikov equation and with a body of classic literature on quantum groups and gauge theories. 

We hope that our analysis will provide the ground for several novel developments in the study of massless modes in $AdS_3$ and in the larger context of AdS/CFT. 

\section{\label{sec:ackn}Acknowledgments}

We owe a tremendous debt of gratitude to J. McOrist, A. Prinsloo and M. Wolf, for illuminating discussions and crucial remarks. We also very much thank O. Ohlsson Sax and B. Stefa\'nski for preliminary discussions about antiferromagnetic excitations in the massless sector of $AdS_3$, and B. Hoare, N. Dorey, M. de Leeuw and A. Pittelli for very useful discussions. A.F. is partially supported by the EPSRC grant FP/M506655. A.T. thanks
the EPSRC for funding under the First Grant project EP/K014412/1 and the the STFC under the Consolidated
Grant project nr. ST/L000490/1.
A.T. also thanks the organisers of the conferences {\em All about $AdS_3$} (ETH, 2015), {\it Selected Topics in Theoretical High Energy Physics} (Tbilisi, 2015) and of the program {\em Holography and Dualities 2016: New Advances in String and Gauge Theory} (Nordita, 2016) for the stimulating atmosphere.

\medskip

{\bf \small Data management:} No data beyond those presented and cited in this work are needed to validate this study.

\appendix
\section{Consistency check}
In this appendix, we verify the path-ordered formula for the $R$-matrix on a specific contour, which is most convenient for the calculation. We shall take a particular straight line, such that (\ref{7}) is integrated to
\begin{equation}
\label{Wilsone}
R = \Pi_s \, {\cal{P}} e^{\int_{p_2}^{p_1} dx \Gamma_1 (x,p_2)}. 
\end{equation} 
We begin by expanding out (\ref{Wilsone}) as follow
\begin{eqnarray}
&&{\cal{P}} e^{\int_{p_2}^{p_1} dx \, \Gamma_1 (x,p_2)} = \mathfrak{1} + \int_{p_2}^{p_1} dx \, \Gamma_1 (x,p_2) \nonumber\\
&&\qquad + \int_{p_2}^{p_1} dx \int_{p_2}^x dy \, \Gamma_1 (x,p_2) \, \Gamma_1(y,p_2) + ...,
\end{eqnarray}
and, by using (\ref{first}) for $\Gamma_1$, we obtain
\begin{eqnarray}
\label{retu}
&&R = E_{11} \otimes E_{11} - E_{22} \otimes E_{22} +\nonumber\\
&&\qquad \big(E_{11} \otimes E_{22} - E_{22} \otimes E_{11}\big) \, \sin \int_{p_2}^{p_1} g(x,p_2) \, dx +\nonumber \\
&&\qquad \big(E_- \otimes E_+ - E_+ \otimes E_-\big) \, \cos \int_{p_2}^{p_1} g(x,p_2) \, dx,
\end{eqnarray}
where $E_{11} \equiv \begin{pmatrix}1&0\\0&0\end{pmatrix}$, $E_{22} \equiv \begin{pmatrix}0&0\\0&1\end{pmatrix}$, and
\begin{equation}
\label{g}
g(p_1,p_2) \equiv -\frac{1}{4} \sqrt{\frac{\sin \frac{p_2}{2}}{\sin \frac{p_1}{2}}} \, \frac{1}{\sin \frac{p_1 + p_2}{4}}. 
\end{equation}
This can be explicitly verified to coincide with (\ref{eq:RLL}), remembering that $\mathfrak{a} \otimes \mathfrak{b} \, |v\rangle \otimes |w\rangle = (-)^{|\mathfrak{b}||v|} \mathfrak{a}|v\rangle \otimes \mathfrak{b}|w\rangle$.

Because $\Gamma_M = g_M (p_1,p_2) \big[E_+ \otimes E_- + E_- \otimes E_+\big]$, with $g_1 (p_1,p_2)= g(p_1,p_2)$ in (\ref{g}) and $g_2 (p_1,p_2) = - g_1 (p_2,p_1)$, clearly the choice 
\begin{equation}
\label{choice}
\frac{dp^M}{d\tau} = \frac{1}{g_M\big(p^1(\tau),p^2(\tau)\big)}
\end{equation}
reduces the auxiliary Hamiltonian to the constant matrix $E_+ \otimes E_- + E_- \otimes E_+$. Integrating (\ref{choice}) with the specified boundary conditions gives an alternative proof of (\ref{retu}).

In terms of the scattering coefficients, we then write
\begin{equation}
R = \begin{pmatrix}1&0&0&0\\0&\sin \sigma &\cos \sigma&0\\0&\cos \sigma&-\sin \sigma&0\\0&0&0&-1\end{pmatrix}, \qquad \sigma \equiv \int _{p_2}^{p_1} g(x,p_2) \, dx.\nonumber
\end{equation}
By reordering the states as $\{|\phi\rangle \otimes |\phi\rangle,|\psi\rangle \otimes |\psi\rangle,|\psi\rangle \otimes |\phi\rangle,|\phi\rangle \otimes |\psi\rangle\}$, $R$ can be written as
\begin{equation}
R = \begin{pmatrix}1&0&0&0\\0&-1 &0&0\\0&0&0&1\\0&0&1&0\end{pmatrix}\cdot \begin{pmatrix}1&0&0&0\\0&1 &0&0\\0&0&\cos \sigma & \sin \sigma\\0&0&-\sin \sigma&\cos \sigma\end{pmatrix}.\nonumber
\end{equation}

\section{Alternative interpretations}

\subsubsection{Similarities} 

It is interesting to mention that $\Delta(\mathfrak{J})$ resembles a (deformed) super-Poincar\'e generator in two dimensions, whose typical undeformed version reads in superspace
$
J^{\alpha \beta} = x^\alpha P^\beta - x^\beta P^\alpha - \frac{1}{2} P_\gamma \bar{\theta} \gamma^{\alpha \beta \gamma} \theta.
$
This would present $R$ as a $q$-super-translation invariant. 

\smallskip

Additionally, (\ref{7}) reminds of Knizhnik-Zamolodchikov (KZ) equations and their quantisation \cite{FR}. The analogy with the KZ equation becomes stronger when considering that the ``matrix" part of (\ref{7}) is proportional to $\check{r}$ of \cite{Joakim}. It would be fascinating to connect this to $q$-CFTs \cite{Sara}, or form factors in integrable models \cite{Smirnov:1993gp}. 

\subsubsection{Universal $R$-matrix}
We observe that (\ref{int_R}) could be rewritten in terms of the supercharges $\mathfrak{Q}$ and $\mathfrak{S}$, by recombining suitable factors of $\sqrt{\sin \frac{p}{2}}$ in the exponent. When so expressed, we believe this should provide equivalent rewritings of the {\it universal $R$-matrix} of the $q$-deformed Poincar\'e Hopf-superalgebra. Not surprisingly, universal $R$-matrices are traditionally given by exponential formulas. It would be interesting to verify this claim from first principles in view of \cite{Beisert:2016qei}, and get an algebraic expression for the scalar factor \cite{upcom}.  

When regarded in this perspective, our approach is very reminiscent of the one developed in \cite{Maillet}. Here, the standpoint is slightly different, as in our particular case the ordinary classical $r$-matrix cannot be defined \cite{Joakim}. Nevertheless, the two procedures become very close in appearance when considering $\check{r}$ \cite{Joakim}, with the crucial distinction that the latter is not a solution of the classical Yang-Baxter equation. We feel however that there should be a strong relationship, given the striking resemblance.

\subsubsection{Berry phase}
One might conceive regimes where the geometric and quantum mechanical interpretations we have outlined in the main text converge into a single picture, inspired by the notion of {\it Berry phase}. This might tie in with the link drawn in \cite{Joakim} with the physics of phonons, excitations created by particles moving in the potential of slowly-vibrating ions in a crystal. From the viewpoint of the $q$-Poincar\'e algebra, the momenta $(p_1, p_2)$ cohere as a single phonon \cite{Joakim}, which could be described by a single-particle quantum mechanics. A Berry-phase picture could link to our ${\cal{B}}$ bundle, with the momenta $p^M(\tau)$ as adiabatically-changing parameters (cf. also the so-called {\it vacuum bundles} \cite{Hori}, pointed out to us by J. McOrist).

\subsubsection{Lax pairs}
If we read the flatness of $\Gamma$ in terms of a Lax pair, then this could define a {\it classically integrable} system (although it is not a mathematical implication). If this were the case, this could be yet another subsidiary interpretation. $R$ would then be the solution of the auxiliary linear problem, therefore  connected to the {\it Gel'fand-Levitan- Marchenko} equation (reviewed {\it e.g.} in \cite{revDurham}) giving soliton solutions {\it via} the classical inverse scattering method. One issue is that $[\Gamma_1,\Gamma_2]$ vanishes on its own. Such Lax pairs are sometimes called {\it weak} \cite{colorado} - as the momenta became {\it coordinates}, we have no spectral parameter, and conservation laws trivialise. One could envisage introducing a spectral parameter ({\it baxterisation}). This might affect (or perhaps resolve) some of the singularities of $\Gamma$. It could also provide a link with the recent results of \cite{Klose:2016uur}. We plan to carry out this analysis in a future study.


\begin{thebibliography}{99}

\bibitem{Beisert:2010jr}
N.~Beisert and ~others,
{\emph {Review of AdS/CFT integrability: an overview},}
{Lett.~Math.~Phys. {\bf 99}
  (2012)~3} [{1012.3982
  [hep-th]}].

\bibitem{Arutyunov:2009ga}
G.~Arutyunov and S.~Frolov,
{\em Foundations of the $AdS_5 \times S^5$ superstring. Part I,}
{J.~Phys. A {\bf 42}
  (2009) 254003} [{0901.4937
  [hep-th]}].

  \bibitem{Babichenko:2009dk}
A.~Babichenko, B.~Stefa{\'n}ski, and K.~Zarembo,
{\em Integrability and the {$AdS_3/CFT_2$} correspondence,}
{JHEP {\bf 1003} (2010) 058}
  [{0912.1723 [hep-th]}].

\bibitem{rev3}
  A.~Sfondrini,
  {\em Towards integrability for $AdS_3/CFT_2$,}
  J.\ Phys.\ A {\bf 48} (2015)  023001
  [arXiv:1406.2971 [hep-th]].
  
\bibitem{Sundin:2012gc}
P.~Sundin and L.~Wulff,
{\em Classical integrability and quantum aspects of the {$AdS_3 \times S^3
  \times S^3 \times S^1$} superstring,}
{JHEP {\bf 1210} (2012) 109}
  [{1207.5531 [hep-th]}].

\bibitem{OhlssonSax:2011ms}
O.~O.~Sax and B.~Stefa{\'n}ski,
{\em Integrability, spin-chains, and the {$AdS_3/CFT_2$} correspondence,}
{JHEP {\bf 1108} (2011) 029}
  [{1106.2558 [hep-th]}].

\bibitem{Borsato:2012ud}
R.~Borsato, O.~O.~Sax, and A.~Sfondrini,
{\em {A dynamic $\mathfrak{su}(1|1)^2$ $S$-matrix for $AdS_3/CFT_2$},}
{JHEP {\bf 1304} (2013) 113}
  [{1211.5119 [hep-th]}].

\bibitem{Borsato:2012ss}
R.~Borsato, O.~O.~Sax, and A.~Sfondrini,
{\em {All-loop Bethe ansatz equations for $AdS_3/CFT_2$},}
{JHEP {\bf 1304} (2013) 116}
  [{1212.0505 [hep-th]}].

\bibitem{Borsato:2013qpa}
R.~Borsato, O.~O.~Sax, A.~Sfondrini, B.~Stefa\'nski, and A.~Torrielli,
{\em {The all-loop integrable spin-chain for strings on AdS$_3 \times S^3
  \times T^4$: the massive sector},}
{JHEP {\bf 1308} (2013) 043}
  [{1303.5995}].

\bibitem{Borsato:2013hoa}
  R.~Borsato, O.~Ohlsson Sax, A.~Sfondrini, B.~Stefanski, Jr. and A.~Torrielli,
  {\em Dressing phases of $AdS_3/CFT_2$,}
  Phys.\ Rev.\ D {\bf 88} (2013) 066004
  [arXiv:1306.2512 [hep-th]].

\bibitem{Rughoonauth:2012qd}
N.~Rughoonauth, P.~Sundin, and L.~Wulff,
{\em Near {BMN} dynamics of the {$AdS_3 \times S^3 \times S^3 \times S^1$}
  superstring,}
{JHEP {\bf 1207} (2012) 159}
  [{1204.4742 [hep-th]}].

\bibitem{Abbott:2012dd}
M.~C.~Abbott,
{\em {Comment on strings in $AdS_3 \times S^3 \times S^3 \times S^1$ at one
  loop},}
{JHEP {\bf 1302} (2013) 102}
  [{1211.5587 [hep-th]}].

\bibitem{Beccaria:2012kb}
M.~Beccaria, F.~Levkovich-Maslyuk, G.~Macorini, and A.~Tseytlin,
{\em {Quantum corrections to spinning superstrings in $AdS_3\times S^3 \times
  M^4$: determining the dressing phase},}
{JHEP {\bf 1304} (2013) 006}
  [{1211.6090 [hep-th]}].

\bibitem{Beccaria:2012pm}
M.~Beccaria and G.~Macorini,
{\em {Quantum corrections to short folded superstring in $AdS_3 \times S^3
  \times M^4$},}
{JHEP {\bf 1303} (2013) 040}
  [{1212.5672 [hep-th]}].

\bibitem{Sundin:2013ypa}
P.~Sundin and L.~Wulff,
{\em {World-sheet scattering in $AdS_3/CFT_2$},}
{JHEP {\bf 1307} (2013) 007}
  [{1302.5349 [hep-th]}].

\bibitem{Bianchi:2013nra}
L.~Bianchi, V.~Forini, and B.~Hoare,
{\em {Two-dimensional $S$-matrices from unitarity cuts},}
{JHEP {\bf 1307} (2013) 088}
  [{1304.1798 [hep-th]}].
$\bullet$
O.~T.~Engelund, R.~W.~McKeown and R.~Roiban,
{\em Generalized unitarity and the worldsheet $S$-matrix in $AdS_n \times S^n \times M^{10-2n}$,}
JHEP {\bf 1308} (2013) 023
[arXiv:1304.4281].
$\bullet$
  L.~Bianchi and B.~Hoare,
  {\em $AdS_3 \times S^3 \times M^4$ string $S$-matrices from unitarity cuts,}
  JHEP {\bf 1408} (2014) 097
  [arXiv:1405.7947 [hep-th]].

\bibitem{Sax:2012jv}
O.~O.~Sax, B.~Stefa\'nski, and A.~Torrielli,
{\em {On the massless modes of the $AdS_3/CFT_2$ integrable systems},}
{JHEP {\bf 1303} (2013) 109}
  [{1211.1952 [hep-th]}].

\bibitem{Lloyd:2013wza}
T.~Lloyd and B.~Stefa\'nski,
{\em {$AdS_3/CFT_2$, finite-gap equations and massless modes},}
{JHEP {\bf 1404} (2014) 179}
  [{1312.3268 [hep-th]}].

\bibitem{Borsato:2014exa}
  R.~Borsato, O.~Ohlsson Sax, A.~Sfondrini and B.~Stefanski,
  {\em Towards the All-Loop Worldsheet S Matrix for $AdS_3\times S^3\times T^4$,}
  Phys.\ Rev.\ Lett.\  {\bf 113} (2014) no.13,  131601
  [arXiv:1403.4543 [hep-th]].

\bibitem{Borsato:2014hja}
  R.~Borsato, O.~Ohlsson Sax, A.~Sfondrini and B.~Stefanski,
  {\em The complete AdS$_{3} \times$ S$^3 \times$ T$^4$ worldsheet S matrix,}
  JHEP {\bf 1410} (2014) 66
  [arXiv:1406.0453 [hep-th]].

\bibitem{Abbott:2014rca}
  M.~C.~Abbott and I.~Aniceto,
  {\em Macroscopic (and Microscopic) Massless Modes,}
  Nucl.\ Phys.\ B {\bf 894} (2015) 75
  [arXiv:1412.6380 [hep-th]].
  $\bullet$
  M.~C.~Abbott and I.~Aniceto,
  {\em Massless L\"uscher Terms and the Limitations of the $AdS_3$ Asymptotic Bethe ansatz}
  Phys.\ Rev.\ D {\bf 93} (2016) no.10,  106006
  [arXiv:1512.08761 [hep-th]].

\bibitem{Sax:2014mea}
  O.~O.~Sax, A.~Sfondrini and B.~Stefanski,
  {\em Integrability and the Conformal Field Theory of the Higgs branch,}
  JHEP {\bf 1506} (2015) 103
  [arXiv:1411.3676 [hep-th]].
  
\bibitem{Borsato:2016kbm}
  R.~Borsato, O.~O.~Sax, A.~Sfondrini and B.~Stefanski,
  {\em On the spectrum of $AdS_3 \times S^3 \times T^4$ strings with Ramond-Ramond flux,}
  arXiv:1605.00518 [hep-th].

\bibitem{upcom}
  R.~Borsato, O.~Ohlsson Sax, A.~Sfondrini, B.~Stefanski and A.~Torrielli,
  {\it On the Dressing Factors, Bethe Equations and Yangian Symmetry of Strings on $AdS_3 \times S^3 \times T^4$,}
  arXiv:1607.00914 [hep-th].
  
\bibitem{PerLinus}
  P.~Sundin and L.~Wulff,
  {\em The complete one-loop BMN $S$-matrix in $AdS_3 \times S^3 \times T^4$,}
  arXiv:1605.01632 [hep-th].
  
  \bibitem{Abbott:2013ixa}
M.~C.~Abbott,
{\em {The $AdS_{3} \times S^{3} \times S^{3} \times S^{1}$ Hern\'andez-Lopez
  phases: a semiclassical derivation},}
{J.~Phys. A {\bf 46}
  (2013) 445401} [{1306.5106
  [hep-th]}].

\bibitem{Sundin:2013uca}
P.~Sundin and L.~Wulff,
{\em {The low energy limit of the $AdS_{3} \times S^{3} \times M_{4}$ spinning
  string},}
{JHEP {\bf 1310} (2013) 111}
  [{1306.6918 [hep-th]}].

\bibitem{Borsato:2015mma}
R.~Borsato, O.~Ohlsson Sax, A.~Sfondrini and B.~Stefa\'nski,
  {\em The $AdS_3\times S^3\times S^3 \times S^1$ worldsheet S matrix,}
  J.\ Phys.\ A {\bf 48} (2015)  415401
  [arXiv:1506.00218 [hep-th]].
  
\bibitem{Prin}
A.~Prinsloo,
  {\em D1 and D5-brane giant gravitons on $AdS_3 \times S^3 \times S^3 \times S^1$,}
  JHEP {\bf 1412} (2014) 094
  [arXiv:1406.6134 [hep-th]].
$\bullet$
A.~Prinsloo, V.~Regelskis and A.~Torrielli,
  {\em Integrable open spin-chains in $AdS_3/CFT_2$ correspondences,}
  Phys.\ Rev.\ D {\bf 92} (2015) no.10,  106006
  [arXiv:1505.06767 [hep-th]].
  
\bibitem{Abbott:2015mla}
M. C. Abbott, J. Murugan, S. Penati, A. Pittelli, D. Sorokin, P. Sundin, J. Tarrant, M. Wolf and L. Wulff,
  {\em T-duality of Green-Schwarz superstrings on 
  $AdS_d \times S^d \times M^{10-2d}$,}
  JHEP {\bf 1512} (2015) 104
  [arXiv:1509.07678 [hep-th]].
$\bullet$
M.~C.~Abbott, J.~Tarrant and J.~Murugan,
  {\em Fermionic T-Duality of $AdS_n \times S^n (\times S^n) \times T^m$ using IIA Supergravity,}
  Class.\ Quant.\ Grav.\  {\bf 33} (2016) 075008
  [arXiv:1509.07872 [hep-th]].
  
\bibitem{Per}
J.~R.~David and B.~Sahoo,
  {\em Giant magnons in the D1-D5 system,}
  JHEP {\bf 0807} (2008) 033
  [arXiv:0804.3267 [hep-th]].
$\bullet$
J.~R.~David and B.~Sahoo,
  {\em $S$-matrix for magnons in the D1-D5 system,}
  JHEP {\bf 1010} (2010) 112
  [arXiv:1005.0501 [hep-th]].
$\bullet$
C.~Ahn and D.~Bombardelli,
  {\em Exact $S$-matrices for $AdS_3/CFT_2$,}
  Int.\ J.\ Mod.\ Phys.\ A {\bf 28} (2013) 1350168
  [arXiv:1211.4512 [hep-th]].
$\bullet$
  M.~C.~Abbott,
  {\em The $AdS_{3} \times S^{3} \times S^{3} \times S^{1}$ Hern\'andez-L\'opez phases: a semiclassical derivation,}
  J.\ Phys.\ A {\bf 46} (2013) 445401
  [arXiv:1306.5106 [hep-th]].
    $\bullet$
  P.~Sundin and L.~Wulff,
  {\em The low energy limit of the $AdS_3 \times S^3 \times M^4$ spinning string,}
  JHEP {\bf 1310} (2013) 111
  [arXiv:1306.6918 [hep-th]].
$\bullet$
P.~Sundin,
  {\em Worldsheet two- and four-point functions at one loop in $AdS_3/CFT_2$,}
  Phys.\ Lett.\ B {\bf 733} (2014) 134
  [arXiv:1403.1449 [hep-th]].
  $\bullet$
  R.~Roiban, P.~Sundin, A.~Tseytlin and L.~Wulff,
  {\em The one-loop worldsheet $S$-matrix for the $AdS_n \times S^n \times T^{10-2n}$ superstring,}
  JHEP {\bf 1408} (2014) 160
  [arXiv:1407.7883 [hep-th]].
$\bullet$
M.~C.~Abbott and I.~Aniceto,
  {\em An improved AFS phase for AdS$_3$ string integrability,}
  Phys.\ Lett.\ B {\bf 743} (2015) 61
  [arXiv:1412.6863 [hep-th]].
  $\bullet$
  L.~Wulff,
  {\em On integrability of strings on symmetric spaces,}
  JHEP {\bf 1509} (2015) 115
  [arXiv:1505.03525 [hep-th]].
  $\bullet$
  P.~Sundin and L.~Wulff,
  {\em The $AdS_{n} \times S^{n} \times T^{10-2n}$ BMN string at two loops,}
  JHEP {\bf 1511} (2015) 154
  [arXiv:1508.04313 [hep-th]].

\bibitem{Pittelli:2014ria}
  A.~Pittelli, A.~Torrielli and M.~Wolf,
  {\em Secret symmetries of type IIB superstring theory on $AdS_3 \times S^3 \times M^4$,}
  J.\ Phys.\ A {\bf 47} (2014) no.45,  455402
  [arXiv:1406.2840 [hep-th]].

\bibitem{Regelskis:2015xxa}
  V.~Regelskis,
  {\em Yangian of $AdS_3/CFT_2$ and its deformation,}
  J.\ Geom.\ Phys.\  {\bf 106} (2016) 213
  [arXiv:1503.03799 [math-ph]].

\bibitem{Gomez:2007zr}
  C.~Gomez and R.~Hernandez,
  {\em Quantum deformed magnon kinematics,}  
JHEP {\bf 0703} (2007) 108
  [hep-th/0701200].

\bibitem{Charles}
  C.~A.~S.~Young,
  {\em q-deformed supersymmetry and dynamic magnon representations,}
  J.\ Phys.\ A {\bf 40} (2007) 9165
  [arXiv:0704.2069 [hep-th]].

\bibitem{Pachol:2015mfa}
    A.~Pacho\l{} and S.~J.~van Tongeren,
  {\em Quantum deformations of the flat space superstring,}
  Phys.\ Rev.\ D {\bf 93} (2016) 026008
  [arXiv:1510.02389 [hep-th]].

\bibitem{Kenta}
I.~Kawaguchi and K.~Yoshida,
  {\it Classical integrability of Schrodinger sigma models and q-deformed Poincare symmetry,}
  JHEP {\bf 1111} (2011) 094
  [arXiv:1109.0872 [hep-th]].
  $\bullet$
  I.~Kawaguchi and K.~Yoshida,
  {\it Exotic symmetry and monodromy equivalence in Schrodinger sigma models,}
  JHEP {\bf 1302} (2013) 024
  [arXiv:1209.4147 [hep-th]].
$\bullet$
I.~Kawaguchi, T.~Matsumoto and K.~Yoshida,
  {\it Schroedinger sigma models and Jordanian twists,}
  JHEP {\bf 1308} (2013) 013
  [arXiv:1305.6556 [hep-th]].
    
\bibitem{Joakim}
  J.~Stromwall and A.~Torrielli,
  {\it AdS3/CFT2 and q-Poincare' superalgebras,}
  arXiv:1606.02217 [hep-th].

\bibitem{Ballesteros:1999ew}
  A.~Ballesteros, E.~Celeghini and F.~J.~Herranz,
  {\em Quantum (1+1) extended Galilei algebras: from Lie bialgebras to quantum $R$-matrices and integrable systems,}
  J.\ Phys.\ A {\bf 33} (2000) 3431
  [math/9906094 [math.QA]].
  $\bullet$
  A.~Ballesteros, E.~Celeghini and F.~J.~Herranz,M.~A.~Del~Olmo, M.~Santander,
  {\em Universal $R$-matrices for non-standard (1+1) quantum groups,}
  J.\ Phys.\ A {\bf 28} (1995) 3129.
  $\bullet$
  F.~Bonechi, E.~Celeghini, R.~Giachetti, E.~Sorace and M.~Tarlini,
  {\em Inhomogeneous quantum groups as symmetries of phonons,}
  Phys.\ Rev.\ Lett.\  {\bf 68} (1992) 3718
  [hep-th/9201002].
  $\bullet$
  F.~Bonechi, E.~Celeghini, R.~Giachetti, E.~Sorace and M.~Tarlini,
  {\em Quantum Galilei group as symmetry of magnons,}
  Phys.\ Rev.\ B {\bf 46} (1992) 5727
  [hep-th/9203048].

\bibitem{Kulish}
P.~P.~Kulish and P.~D.~Ryasichenko, {\it Spin chain connected to the quantum superalgebra $\mathfrak{sl}_q(1|1)$,} Zapiski Nauchnykh Seminarov POMI 325 (2005) 146.

\bibitem{Faddeev:1981ip}
  L.~D.~Faddeev and L.~A.~Takhtajan,
  {\it What is the spin of a spin wave?,}
  Phys.\ Lett.\ A {\bf 85} (1981) 375.

\bibitem{H1}
  A.~Rej, D.~Serban and M.~Staudacher,
  {\it Planar N=4 gauge theory and the Hubbard model,}
  JHEP {\bf 0603} (2006) 018
  [hep-th/0512077].
$\bullet$
  G.~Feverati, D.~Fioravanti, P.~Grinza and M.~Rossi,
  {\it Hubbard's Adventures in N = 4 SYM-land? Some non-perturbative considerations on finite length operators,}
  J.\ Stat.\ Mech.\  {\bf 0702} (2007) P02001
  [hep-th/0611186].
  
\bibitem{H2}
  R.~Roiban, A.~Tirziu and A.~A.~Tseytlin,
  {\it ``Slow-string limit and `antiferromagnetic' state in AdS/CFT,}
  Phys.\ Rev.\ D {\bf 73} (2006) 066003
  [hep-th/0601074].
$\bullet$
R.~Ishizeki and M.~Kruczenski,
  {\it Single spike solutions for strings on $S^2$ and $S^3$,}
  Phys.\ Rev.\ D {\bf 76} (2007) 126006
  [arXiv:0705.2429 [hep-th]].
$\bullet$
K.~Okamura,
  {\it Giant Spinons,}
  JHEP {\bf 1004} (2010) 033
  [arXiv:0911.1528 [hep-th]].
  
\bibitem{FR}
  I.~B.~Frenkel and N.~Y.~Reshetikhin,
 {\it Quantum affine algebras and holonomic difference equations,}
  Commun.\ Math.\ Phys.\  {\bf 146} (1992) 1.

\bibitem{Sara}
  F.~Nieri, S.~Pasquetti and F.~Passerini,
  {\it 3d and 5d Gauge Theory Partition Functions as $q$-deformed CFT Correlators,}
  Lett.\ Math.\ Phys.\  {\bf 105} (2015) no.1,  109
  [arXiv:1303.2626 [hep-th]].
$\bullet$  
  F.~Nieri, S.~Pasquetti, F.~Passerini and A.~Torrielli,
  {\it 5D partition functions, q-Virasoro systems and integrable spin-chains,}
  JHEP {\bf 1412} (2014) 040
  [arXiv:1312.1294 [hep-th]].
  
\bibitem{Smirnov:1993gp}
  F.~A.~Smirnov,
  {\it Form-factors, deformed Knizhnik-Zamolodchikov equations and finite gap integration,}
  Commun.\ Math.\ Phys.\  {\bf 155} (1993) 459
  [hep-th/9210052].
  
\bibitem{Beisert:2016qei}
  N.~Beisert, M.~de Leeuw and R.~Hecht,
  {\it Maximally extended $\mathfrak{sl}(2|2)$ as a quantum double,}
  arXiv:1602.04988 [math-ph].
  
\bibitem{Maillet}
  E.~Witten,
  {\it Gauge Theories, Vertex Models and Quantum Groups,}
  Nucl.\ Phys.\ B {\bf 330} (1990) 285.
  $\bullet$
J.~M.~Maillet,
  {\it Integrable systems and gauge theories,}
  Nucl.\ Phys.\ Proc.\ Suppl.\  {\bf 18B} (1991) 212.
$\bullet$
 L.~Freidel and J.~M.~Maillet,
  {\it The Universal R matrix and its associated quantum algebra as functionals of the classical r matrix: The $\mathfrak{sl}(2)$ case,}
  Phys.\ Lett.\ B {\bf 296} (1992) 353
  [hep-th/9210039].
  
\bibitem{Hori}
  K.~Hori, S.~Katz, A.~Klemm, R.~Pandharipande, R.~Thomas, C.~Vafa, R.~Vakil and E.~Zaslow,
  {\it Mirror symmetry,} 
    Clay mathematics monographs, 1
    Providence, USA: AMS (2003).

\bibitem{revDurham}
  A.~Torrielli,
  {\it Lectures on Classical Integrability,}
  J.\ Phys.\ A {\bf 49} (2016) no.32,  323001
  [arXiv:1606.02946 [hep-th]], A. Cagnazzo, R. Frassek, A. Sfondrini,
I. M. Sz\'ecs\'enyi, S. J. van Tongeren editors.


\bibitem{colorado}
J.~Rezac, 
{\it Computation of Scaling Invariant Lax Pairs with Applications to Conservation Laws,} Diss. Colorado School of Mines, 2012.

\bibitem{Klose:2016uur}
  T.~Klose, F.~Loebbert and H.~Munkler,
  {\it Master Symmetry for Holographic Wilson Loops,}
  arXiv:1606.04104 [hep-th].
  
          \end{thebibliography}
\end{document}